\documentclass[aps,prl,10pt,twocolumn,showpacs,superscriptaddress,nofootinbib]{revtex4-1}
\usepackage{graphicx}
\usepackage{verbatim}
\usepackage[usenames,dvipsnames,svgnames]{pstricks}
\usepackage{epsfig}
\usepackage{pst-grad} 
\usepackage{pst-plot} 
\usepackage{epstopdf}
\usepackage{newlfont}
\usepackage{amssymb}
\usepackage{amsfonts}
\usepackage{amsmath}
\usepackage{amsthm}
\usepackage{dsfont}
\usepackage{epsfig}
\usepackage{color}
\usepackage{bm}
\usepackage[breaklinks]{hyperref}
\usepackage{tikz}

\usepackage{times}

\begin{document}
\title{Affect of Simulated Atmospheric Turbulence on Three Entangled Photons}
\affiliation{Cryptology and Security Research Unit, R.C. Bose Center for Cryptology and Security, Indian Statistical Institute, Kolkata 700108, India}
\author{Pritam Chattopadhyay, Ayan Mitra, Goutam Paul, Ram Soorat}
\email{pritam.cphys@gmail.com, amitra@lpnhe.in2p3.fr, goutam.paul@isical.ac.in, rsoorat@gmail.com}

\begin{abstract}
 The thorough experimental study for the measure of sustainability of the atmospheric turbulence of the entangled photons for two qubit has been analysed by G. Puentes et al. [Phys. Rev. A 75, 032319 (2007)]. Here in this paper we propose a scheme to measure the effect of entanglement in  three entangled photon GHZ state with the help of simulated atmospheric turbulence. We simulate the result of affected entangled GHZ state by introducing turbulence in three entangled photon path alternatively. We observe, by introducing the turbulence in photon path we are loosing the entanglement property, because of  the increase in entropy. We compare our result with Werner curve and observe that the simulated result perfectly fit with the theoretical result.
\end{abstract}

\maketitle
\section{Introduction}
\textcolor{black}{The emergence of quantum information theory made quantum entanglement an interesting subject to study. It has been the backbone for various quantum protocols}.  \textcolor{black}{Quantum entanglement has various applications in the field of information theory. It is generally utilized to study quantum logical gates \cite{nielsen, raussendrof},} to \textcolor{black}{carry-out} computation \textcolor{black}{of} algorithms \textcolor{black}{more efficiently}~\cite{montanaro, terhal}\textcolor{black}{.} \textcolor{black}{Even it is used for sharing} information \textcolor{black}{with more security}  \cite{gisin, scarani} in quantum computers and networks \cite{kimble, ladd, van}\textcolor{black}{.} \textcolor{black}{Hyper-dense coding was the first application\cite{mattle, kwiat} that was experimentally studied.} \textcolor{black}{In the resent past, the bipartite entangled state has been explored immensely, whereas the development of multipartite entangled states is quite challenging }. Recently, \textcolor{black}{due to the development of} quantum state fusion \cite{ zang3, lik} and expansion \cite{ tashima1, zang5, yesilyurt1} technology\textcolor{black}{,} large scale \textcolor{black}{of} multipartite entangled states \textcolor{black}{can be created.} \textcolor{black}{It has} attracted much attention \textcolor{black}{in recent times}. 

\textcolor{black}{After the through exploration of two-photon entanglement we encounter a enormous demand for the production of multi-photon entangled states such as} Greenberger-Horne-Zeilinger (GHZ) states \cite{greenberger}. \textcolor{black}{Till date, spontaneous parametric downconversion (SPDC) is the most prevailing method by which we can produce photonic entangled states}. \textcolor{black}{In the process of producing entangled photons using SPDC, we realize that it produces photons in pairs which causes them to entangle through various degrees of freedom \cite{edamatsu}}. \textcolor{black}{As far as} the best quantum channels \textcolor{black}{are of concern, we can consider GHZ state for this purpose.} \textcolor{black}{It can be used} for teleportation \cite{ deny, zhao}, quantum key distribution \cite{kempe}, and many \textcolor{black}{other} application. \textcolor{black}{To fabricate} multiparticle entangled GHZ state, \textcolor{black}{various methods} have been proposed, such as cavity QED system \cite{guo, behzadi},  optical system \cite{ zang}, ion trap system \cite{duan} and quantum dot system \cite{ luo, xu}.

    \textcolor{black}{In the work \cite{lodahl}, they have studied the } effect of multiple scattering \textcolor{black}{for} single-photon. \textcolor{black}{In the work\cite{ursin, aspelmeyer}, they have verified that the anticipation of maintaining correlation for arbitary distance is authentic.} \textcolor{black}{They have performed} free-space propagation \textcolor{black}{with polarized entangled photons} over distances of up to 144 km.  The robustness of \textcolor{black}{polarized entangled states can be proved by implementing}  scattering processes \textcolor{black}{like} \cite{altewischer}, entangled mixed-state generation by twin-photon scattering \cite{puentes1}, effect of polarization entanglement in photon-photon scattering \cite{dennis}. \textcolor{black}{Now we know that polarized entangled photons are not perturbed by the scattering process as long as they are linear (that can be described by a scattering matrix)} and as long  as \textcolor{black}{we detect the} photons  in a single spatial mode \cite{aiello, van1}. Therefore, \textcolor{black}{we would like} to characterize entanglement decay \textcolor{black}{by using} linear scattering processes  \textcolor{black}{along} with multi-mode detection. 
    
    In this \textcolor{black}{paper} we investigate  the impact \textcolor{black}{of the turbulence} on the \textcolor{black}{polarized entangled} GHZ \textcolor{black}{state} by creating turbulence in one arm. Then we have created the turbulence in two  arms of GHZ state in all combinations. We finally crated  the turbulence in all the three arms of the GHZ state of the generated photons. The experimental and simulation showed the same output.

\section{Experimental Set Up}

\begin{figure}[h!]
  \includegraphics[width=.5\textwidth]{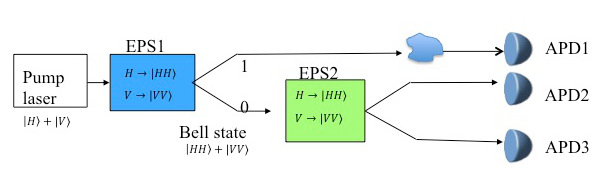}
   \caption{Experimental setup}
  \label{fig1}
\end{figure}

A schematic of the source of direct generation of three entangled photon  based on \textcolor{black}{the work}\cite{ deny} is shown in Fig \ref{fig1}.  A  laser  diode (404 nm)  is  used  to  pump the  first entangled photon  source  (EPS1) which   produces  entangled  photons.  \textcolor{black}{To} produce \textcolor{black}{the} entangled photon pairs \textcolor{black}{we have to tune the temperature when there is a phase matching in the crystal that produces this photon pairs.} \textcolor{black}{The produced photons are} approximately  776 nm and 842 nm. \textcolor{black}{We pump EPS2 which acts as source for second entangled photon by 776 nm photons}, \textcolor{black}{transferring} the entanglement \textcolor{black}{property} to two new photons to produce a GHZ state.  The EPS2 \textcolor{black}{similar to EPS1} are also temperature controlled, \textcolor{black}{and they are} phase matched  to  produce  photons  centered nearly 1530 nm and 1570 nm. The photons 842nm  at 1530 nm and 1570 nm are measured by the  analysers APD1, APD2 and APD3 respectively. \textcolor{black}{To measure} the combined \textcolor{black}{effect of} coupling and detection efficiency of the 842 nm photons, 1530 nm and 1570 nm photons \textcolor{black}{we have to measure} the ratio of photon detections to coincident photon detections. 

  For  this  experiment \textcolor{black}{they have used two types of detectors}. \textcolor{black}{Here in this experiment, they detect} the 842 nm photons  with  silicon avalanche photodiodes (Si-APD). The photons at 1530 nm and 1570 nm are detected by  tungsten silicide superconducting nanowire single-photon detectors  (SNSPD). 

Here in Fig.\ref{fig1} we have triggered the turbulence in the first arm of the setup. We put up the turbulence in all combination to see the difference in the result. Similarly we triggered the turbulence in two of the arms (in all combinations) and finally in all the three arms of the generated GHZ state.

\section{Results}
\subsection{Mathematical analysis for entanglement measure}

 There  are  many methods  available for the measurement of quantum entanglement.  For instance, one could quantify logarithmic negativity \cite{gvid}, entanglement  distillation\cite{pg},  concurrence\cite{run}. We have approached the concurrence method in this paper. We have preferred this method as it is based on the measure of entropy which is the standard method to analyse the chaos in the system.

An important question \textcolor{black}{which is interesting for} both practical and theoretical \textcolor{black}{aspect} is how much noise ad-mixture pure-state entanglement can sustain. The experimental verification for the two photon entanglement is well established \cite{puentes,puentes2}.

The Werner state for the two qubit system as
\begin{equation}\label{a1}
\rho_{ws}(p)=p |\phi\rangle\langle \phi|+\frac{1-p}{4}  \mathds{1}_4,
\end{equation}
where $\phi$ are the Bell State.

 We are developing the three photons system using the same principle as proposed by \cite{puentes}. For three qubit state we consider the Greenberger-Horne-Zeilinger (GHZ) states. The Werner state for the three qubit system considering the GHZ state is defined as:
\begin{equation}\label{a}
\rho_{ws}(p)=p |GHZ\rangle\langle GHZ|+\frac{1-p}{8}  \mathds{1}_8,
\end{equation} 
Where $|GHZ\rangle=\frac{1}{\sqrt{2}} (|000\rangle+|111\rangle)$ and $p$ $\epsilon$ $[0,1]$. 

 So for a $N$ qubit the mathematical form for the Werner state can be expressed as\cite{jens},
\begin{equation}\label{b}
\rho_{ws}(p)=p |\phi_{ME}\rangle\langle \phi_{ME}|+\frac{1-p}{2^N}  \mathds{1}_{2^N}.
\end{equation}
For $N=2$ the state $\phi_{ME}$ represents the Bell states.

Currently we can measure the degree of entanglement by various method like the entanglement distillation \cite{bennett}, the relative entropy of entanglement \cite{vedral}. Here we will measure the entanglement by using linear entropy \cite{bose} which is given by 

\begin{equation}\label{c}
S_L=\frac{4}{3}\{1-Tr[\rho^2]\},
\end{equation}
which range from $0$ (for pure state) to $1$ (for a maximally-mixed state). Here $\rho$ is the density state of the system that is under study. Tangle which is just the square of concurrence \cite{coffman} is given by
\begin{equation}\label{d}
\tau =\textit{C}^2=[max\{\lambda_1- \lambda_2 -\lambda_3 -\lambda_4,0\}]^2.
\end{equation}
Here the $\lambda's$ are the square roots of the eigenvalue of the matrix $\rho \tilde{\rho}=\rho \sigma_y\otimes \sigma_y \rho^* \sigma_y\otimes \sigma_y$ where $\sigma_y$ is the Pauli matrix and is denoted as $ \sigma_y = \left[ \begin{array}{cc} 0 & -i \\ i & 0\\ \end{array}\right]$. For a maximally-entangled pure state $\tau=1$, while we have $\tau=0$ for non-entangled state.

The generic state of the three-qubit system(i.e, A, B and C) in the standard basis, where each index takes the values 0 and 1 is defined as
\begin{equation}
|\eta\rangle = \sum_{i,j,k} a_{ijk} |ijk\rangle,
\end{equation}
where $a_{ijk}$ are the co-efficients of the states and $|ijk\rangle = |i_A\rangle \otimes |j_B\rangle \otimes |k_C\rangle$.

The tangle for three qubit i.e. tangle of A with BC is the sum of tangle  AB, AC, and the three-way tangle. It's mathematical representation stands as\cite{valerie},
\begin{equation}\label{e}
\tau_{A(BC)}=\tau_{AB}+\tau_{Ac}+\tau_{AB}+\tau_{ABC}.
\end{equation}
Here $\tau_{ABC}=4|d_1-2d_2+4d_3|$, where 

\begin{flalign}\label{f}
d_1 = & ~a_{000}^2 a_{111}^2+a_{001}^2 a_{110}^2+a_{010}^2 a_{101}^2+a_{100}^2 a_{011}^2,\\ \nonumber
d_2 = & ~a_{000} a_{111} a_{011} a_{100}+ a_{000} a_{111} a_{101} a_{010} \nonumber\\
 & +a_{000} a_{111} a_{110} a_{001}+a_{011} a_{100} a_{101} a_{010} \nonumber \\
 & +a_{011} a_{100} a_{110} a_{001} + a_{101} a_{010} a_{110} a_{001}, \nonumber
\\
d_3 = & ~a_{000} a_{110} a_{101} a_{011}+a_{111} a_{001} a_{010} a_{100}. \nonumber
\end{flalign}

We can calculate $\tau_{AB}$ and $\tau_{AC}$ by using Eq.~\eqref{d}.  

We want to measure the endurance of the entanglement due to the effect of the turbulence. For turbulence or noise, we have considered atmospheric turbulence which affects the polarization of the entangled photons that are generated by using BBO crystals. The mathematical form of the noise \cite{simon} is given by
\begin{equation}\label{A}
A = \left[
\begin{array}{cc}
cosh\theta & sinh\theta \\
sinh\theta & cosh\theta \\
\end{array}\right],
\end{equation}

 Where $\theta$ $\epsilon$ $[0,\pi]$. We have considered Eq.~\eqref{A} as a noise which affects the polarization of the generated photon. This noise is the general form of the noise which counts all the angles that changes the polarization of the photons \cite{simon}.

It is well known from the theoretical perspective that if entropy increase due to the atmospheric turbulence the entanglement of the photons reduces \cite{puentes}. In this paper we are going to generate entangled GHZ state using three photon. Here we have done a simulation of the experimental set up. 

The photons that are generated from the experimental set up as shown in Fig. \ref{fig1} are polarized entanglement photon. We preferred this type of atmospheric turbulence (Eq.~\eqref{A}) as this affects the polarization property of the entangled photon.

\subsection{Simulation results}

For the computational purpose we considered the density state as $\rho =\frac{1}{2}|GHZ\rangle\langle GHZ|$. We apply turbulence first at one photon path and measure how the atmospheric turbulence affect the entanglement of the generated photon. The density state after applying the noise stands as
\begin{flalign}\label{x} 
\rho = & ~|0_1\rangle |0_2\rangle |0_3\rangle \langle 0_1| \langle 0_2| \langle 0_3| A_1 \nonumber\\
& + |0_1\rangle |0_2\rangle |0_3\rangle \langle 1_1| \langle 1_2| \langle 1_3| A_1 \nonumber\\
& + |1_1\rangle |1_2\rangle |1_3\rangle \langle 0_1| \langle 0_2| \langle 0_3| A_1 \\  \nonumber
 & ~+ |1_1\rangle |1_2\rangle |1_3\rangle \langle 1_1| \langle 1_2| \langle 1_3| A_1.
\end{flalign}

Here the noise is placed in the first photon path which is denoted by $A_1$. Here  $A_1$ is the turbulence defined as in Eq.~\eqref{A}. Similarly we applied the noise individually to the second photon path and on the then on the third photon path. The outcome of the effect in all the cases are the same.

Similarly we  apply turbulence to the two of the photon path in all permutation and the outcome for noise at two photon are always the same which is shown in Fig. ~\ref{fig4} by the delta scattered plot. The mathematical form for the density state when turbulence is applied at two photon path is
\begin{flalign}\label{x1}
\rho = & ~|0_1\rangle |0_2\rangle |0_3\rangle \langle 0_1| \langle 0_2| \langle 0_3| A_1 A_2 \nonumber\\
& + |0_1\rangle |0_2\rangle |0_3\rangle \langle 1_1| \langle 1_2| \langle 1_3| A_1 A_2 \nonumber\\
& + |1_1\rangle |1_2\rangle |1_3\rangle \langle 0_1| \langle 0_2| \langle 0_3| A_1  A_2 \nonumber\\
 & ~+ |1_1\rangle |1_2\rangle |1_3\rangle \langle 1_1| \langle 1_2| \langle 1_3| A_1 A_2.
\end{flalign}
Here $A_1$ and $A_2$ is applied to the first photon path and second photon path. We can develop all the permutation similarly as described in Eq. ~\eqref{x1}.

Following the same procedure  we apply the noise to all the three photon path to measure how turbulence affect the entanglement.The simulation result of this case is shown in Fig. ~\ref{fig4} by the circle scattered plot. The mathematical form for the density state when turbulence is applied at two photon path is
\begin{flalign}\label{x3} 
\rho = & ~|0_1\rangle |0_2\rangle |0_3\rangle \langle 0_1| \langle 0_2| \langle 0_3| A_1 A_2 A_3 \nonumber\\
& + |0_1\rangle |0_2\rangle |0_3\rangle \langle 1_1| \langle 1_2| \langle 1_3| A_1 A_2 A_3 \nonumber \\
& + |1_1\rangle |1_2\rangle |1_3\rangle \langle 0_1| \langle 0_2| \langle 0_3| A_1  A_2 A_3 \nonumber\\
       & ~+ |1_1\rangle |1_2\rangle |1_3\rangle \langle 1_1| \langle 1_2| \langle 1_3| A_1 A_2 A_3.
\end{flalign} 
Here $A_1$, $A_2$, $A_3$ is applied to the first photon, second photon and the third photon path, where $A_1 = A_2 = A_3 = A$.

We applied the turbulence sequentially to all the photon path and in other case we placed the turbulence at  all the photon path at the same instance. For the computation perspective this two situation shows the 
same effect and experimentally we will get same result for both the cases keeping aside all other external effects due to the atmosphere.

\begin{figure}[!]
\begin{center}
\includegraphics[width=0.5\textwidth]{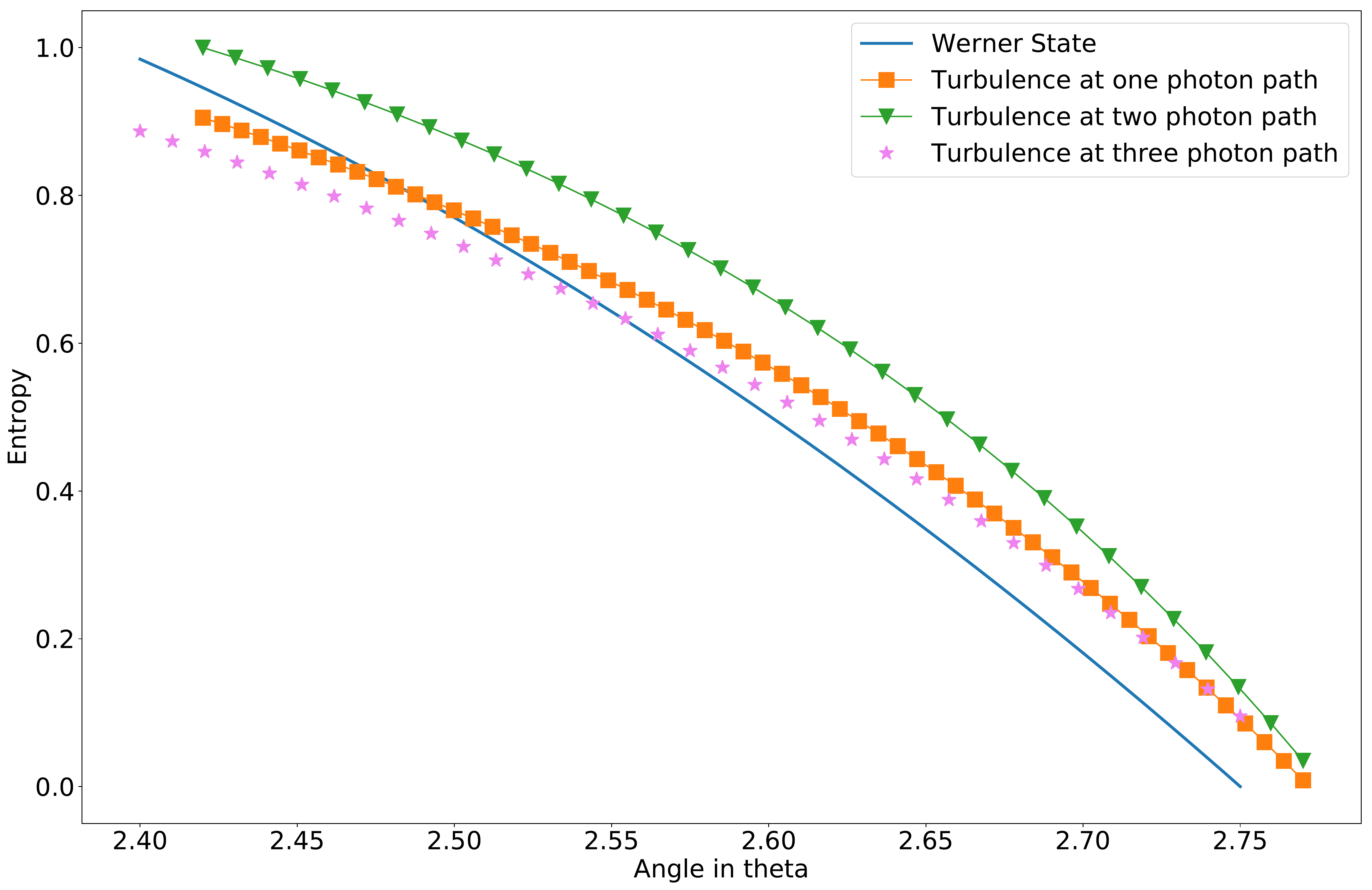}
\end{center}
  \caption{Plot of entropy versus turbulence parameter  for the GHZ state that are generated by photon scattering. The blue line represent the theoretical Werner State (Eq.~\eqref{a}). The pink scattered plot is the simulation of the entangled photon with noise at three of the photon path. Similarly, the green is for noise at two photon path and the orange for the noise at one photon path.}
  \label{fig2}
\end{figure}
\begin{figure}[!]
\begin{center}
\includegraphics[width=0.5\textwidth]{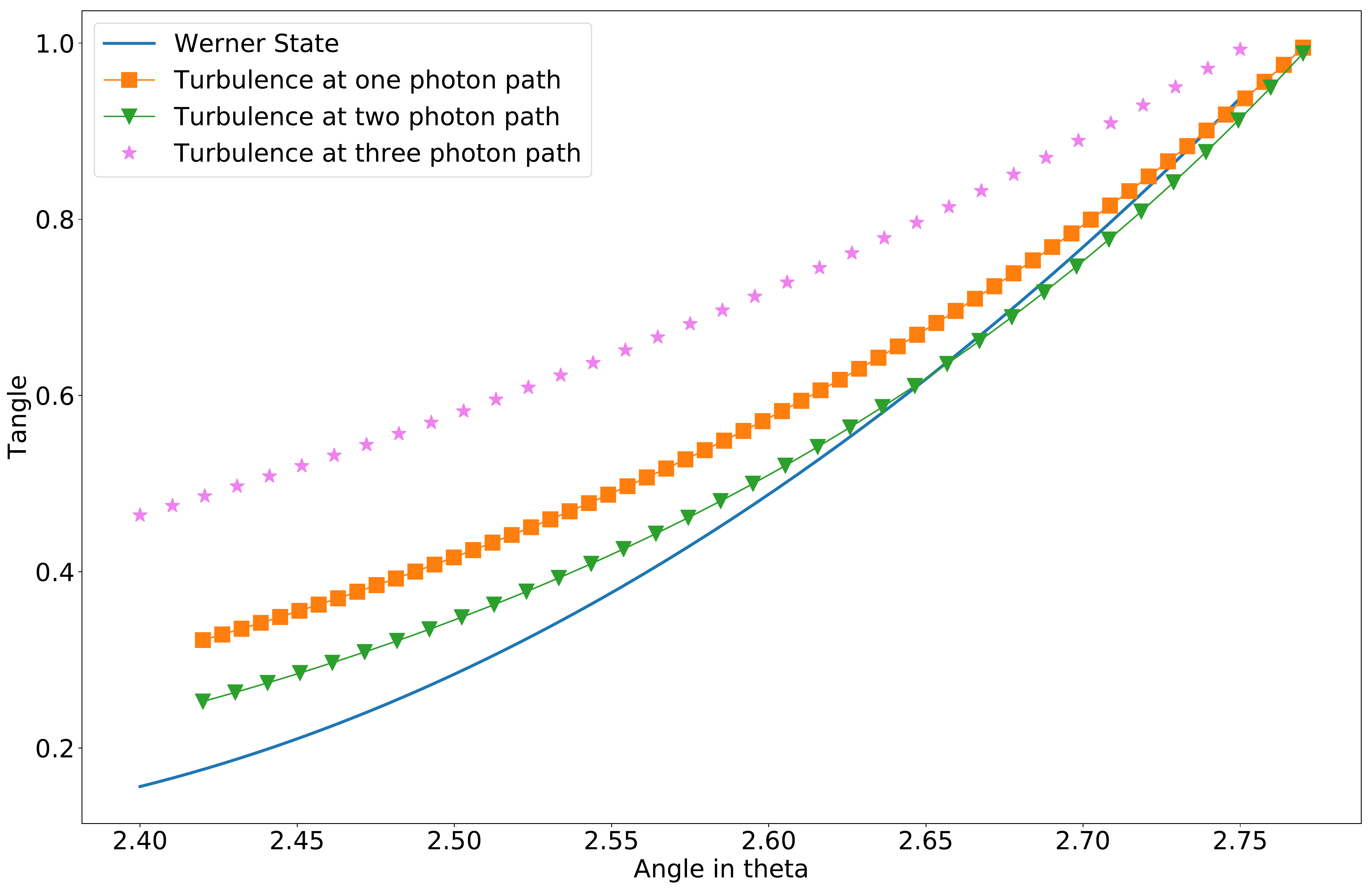}
\end{center}
  \caption{Plot of tangle versus turbulence parameter  for the GHZ state that are generated by photon scattering. The blue line represent the theoretical Werner State (Eq.~\eqref{a}). The pink scattered plot is the simulation of the entangled photon with noise at three of the photon path. Similarly, the green is for noise at two photon path and the orange for the noise at one photon path.}
  \label{fig3}
\end{figure}
 From Fig. ~\ref{fig2} we could visualise the change in the measure of entropy due to the introduction of turbulence in the different combinations to the different photon paths. We have also compared the experimental analysis with our theoretical werner state of three photon. Similarly from Fig. ~\ref{fig3} we can infer the change that is incorporated by the turbulence parameter in the measure of tangle. We have compared our analysis with the theoretical werner state of the three photon generation by scattering process.
\begin{figure}[!]
\begin{center}
\includegraphics[width=0.5\textwidth]{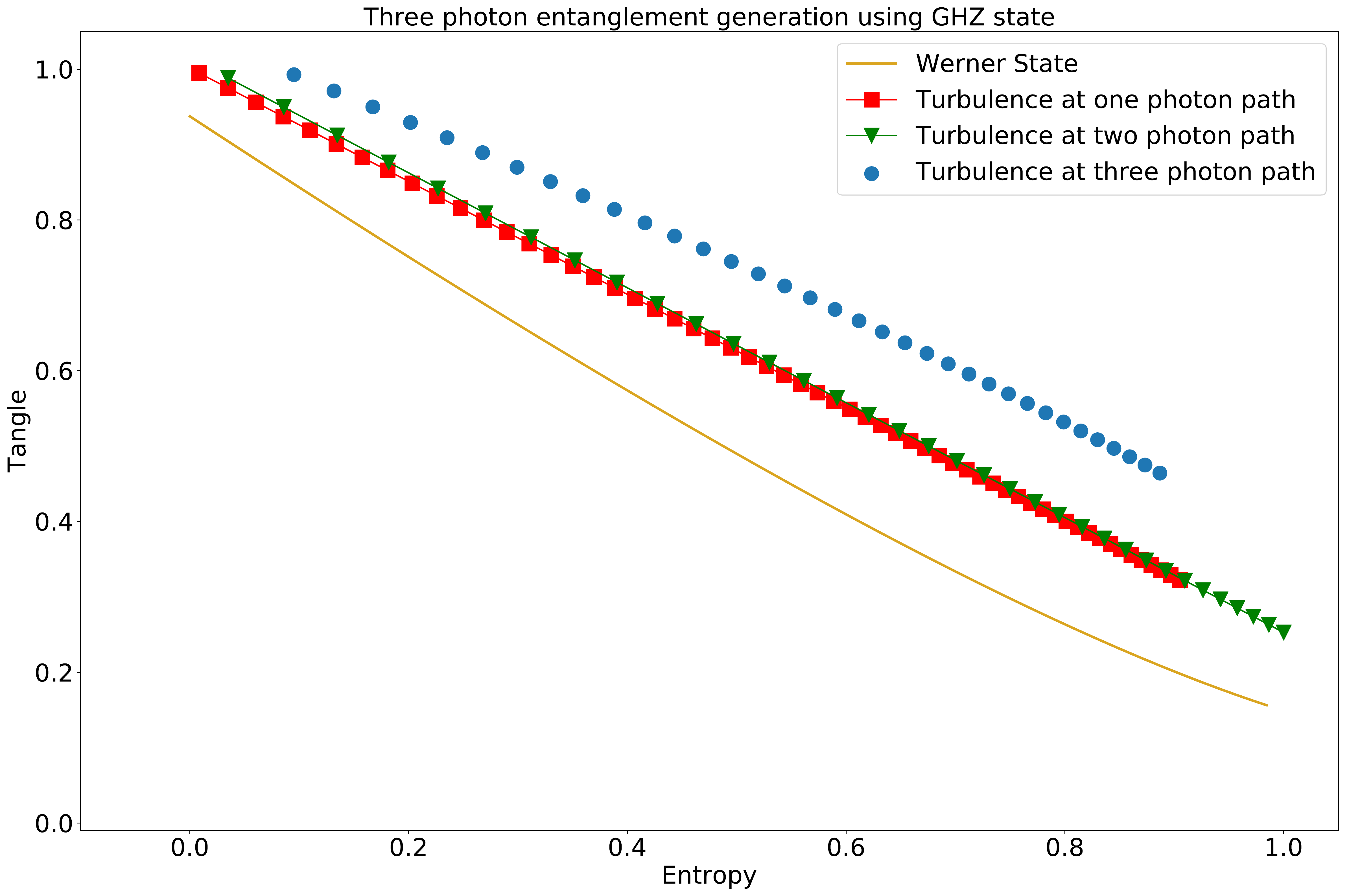}
\end{center}
  \caption{Plot of entropy versus tangle for the GHZ state that are generated by photon scattering. The yellow line represent the theoretical Werner State (Eq.~\eqref{a}). The pink scattered plot is the simulation of the entangled photon with noise at one of the photon path. Similarly, the green is for noise at two photon path and the blue for the noise at three photon path.}
  \label{fig4}
\end{figure}

 The turbulence introduce the parameter which changes the  polarisation of the generated photons. From Fig. ~\ref{fig4} we can infer that the introduction of the noise gives a lateral shift from the Werner curve. The shift is small when the noise is introduced at one of the photon's path(represented by the square scattered plot) and it increases as the noise is included in the two photon paths and then to the three photon path which are represented by the green and blue scattered plot in Fig.~\ref{fig4}. From the plot we can infer that as we insert turbulence in photon path we loose the entanglement property and it increases as we insert it all the photon path.

\section{Method}
A schematic representation of the source of direct generation of three entangled photon  is shown in Fig \ref{fig1}. we have triggered the turbulence in the first arm of the setup then following the same procedure we have triggered the turbulence on second arm and finally on the third arm. Similarly, we triggered the turbulence at two arms simultaneously. Finally we triggered turbulence in all the three arms.

We have made a numerical analysis using the state as shown in Eq. ~\eqref{x}, where we have simulated the different turbulence scenarios, discussed above, by introducing a turbulence matrix in the photon path in each case. Here we have considered those turbulence which affects the polarization of the entangled photon which is shown in Eq. ~\eqref{A}. For turbulence in two photon we have considered the state as Eq. ~\eqref{x1} and for turbulence in three photon path it is shown in Eq. ~\eqref{x3}.

\section{Discussion and Conclusion}
To summarize our results, we see from the plot, that due to the turbulence there is a loss in entanglement property. From the Fig. \ref{fig2} we can infer that as we inject the turbulence in one photon path in all subsequent permutations there is a lateral shift from the theoretical Werner curve. Next, we insert the turbulence to the two photon path in all permutation. Similar to the result that we obtained in the previous case (i.e turbulence in one photon path) we get a lateral shift from Werner curve but more than that of the previous situation. For the turbulence in three photon path the shift was even more than the second case where turbulence was present in two of the photon path. 

From the result we infer that as we insert turbulence in photon path, the disorder in the system increases which reduces the entanglement property between the photons. It increase sequentially if we insert the turbulence in two and three photon path respectively.

This work has a wide range of application in the field of quantum information. Generation of three photon entangled state ensure us that it has a  use  for multi-party secure communication, further we will use this scheme for  development of quantum cheque.



\begin{thebibliography}{99}
\bibitem{mattle} Mattle K et al 1996 Dense coding in experimental quantum communication Phys. Rev. Lett. 76 46569.
\bibitem{kwiat} Kwiat P et al 1996 Polarization-entangled photons and quantum dense coding Opt. Photon. News 7 145.
\bibitem{nielsen}  M. A. Nielsen and I. L. Chuang, Quantum Computation and Quantum Information (Cambridge University, Cambridge, England, 2000).
\bibitem{raussendrof} R. Raussendorf and H. J. Briegel, A One-Way Quantum Computer, Phys. Rev. Lett. 86, 5188 (2001).
\bibitem{montanaro} A. Montanaro, Quantum algorithms: An overview, NPJ Quantum Inf. 2, 15023 (2016).
\bibitem{terhal} B. M. Terhal, Quantum error correction for quantum memories, Rev. Mod. Phys. 87, 307 (2015).
\bibitem{gisin} N. Gisin and R. Thew, Quantum communication, Nat. Photonics  1, 165 (2007).
\bibitem{scarani}  V. Scarani, H. Bechmann-Pasquinucci, N. J. Cerf, M. Duek, N. Lütkenhaus, and M. Peev, The security of practical quantum key distribution, Rev. Mod. Phys. 81, 1301 (2009).
\bibitem{kimble} H. J. Kimble, The quantum internet, Nature (London) 453, 1023 (2008).
\bibitem{ladd} Ladd, T. D., Jelezko, F., Laflamme, R., Nakamura, Y., Monroe, C., \& O’Brien, J. L. (2010). Quantum computers. Nature, 464(7285), 45.
\bibitem{van}  R. Van Meter, Quantum Networking (Wiley, New York, 2014).
\bibitem{zang3} Zang X P, Yang M, Song W and Cao Z L 2016 Fusion of entangled coherent W and GHZ states in cavity QED Opt. Commun. 370 168.
\bibitem{lik}  Li K, Kong F Z, Yang M, Yang Q and Cao Z L, Qubitloss-free fusion of W states Phys. Rev. A 94 062315,2016. 

\bibitem{tashima1}  Tashima T, Özdemir S K, Yamamoto T, Koashi M and Imoto N 2009 Local expansion of photonic Wstate using a polarization- dependent beamsplitter New J. Phys. 11 023024.
\bibitem{zang5}  Zang X P, Yang M, Ozaydin F, Song W and Cao Z L 2016 Deterministic generation of large scale atomic W states Opt. Express 24 12293.
\bibitem{yesilyurt1} Yesilyurt C, Bugu S, Ozaydin F, Altintas A A, Tame M, Yang L and Özdemir S K 2016 Deterministic local expansion of W states J. Opt. Soc. Am. B 33 2313.

\bibitem{greenberger}  Greenberger, D. M., Horne, M. A., Shimony, A., \& Zeilinger, A. (1990). Bell’s theorem without inequalities. American Journal of Physics, 58(12), 1131-1143.
\bibitem{edamatsu} Edamatsu, Keiichi. ``Entangled photons: generation, observation, and characterization." Japanese Journal of Applied Physics 46.11R (2007): 7175..
\bibitem{deny} Deny R. Hamel, Lynden K. Shalm, Hannes Hübel, Aaron J. Miller, Francesco Marsili, Varun B. Verma, Richard P. Mirin, SaeWoo Nam, Kevin J. Resch and Thomas Jennewein ``Direct generation of three-photon polarization entanglement", NATURE PHOTONICS | VOL 8 | OCTOBER 2014 .
\bibitem{zhao} Zhao Z et al 2004 Experimental demonstration of five-photon entanglement and open-destination teleportation Nature 430 54.
\bibitem{kempe} Kempe J 1999 Multiparticle entanglement and its applications to cryptography Phys. Rev. A 60 910.
\bibitem{guo} Guo G P, Li C F, Li J and Guo G C 2002 Scheme for the preparation of multiparticle entanglement in cavity QED Phys. Rev. A 65 042102.
\bibitem{behzadi} Behzadi, Naghi, B. Ahansaz, and S. Kazemi. ``Constructing Robust Entangled Coherent GHZ and W States via a Cavity QED System." International Journal of Theoretical Physics 55.3 (2016): 1577-1592.

\bibitem{zang} Zang X P, Yang M, Wu W F and Fan H Y, Generating multi-mode entangled coherent W and GHZ states via optical system based fusion mechanism Quantum Inf. Process. 16 135, 2017.
\bibitem{duan} Duan L M and Monroe C 2010 Colloquium: quantum networks with trapped ions Rev. Mod. Phys. 82 1209.

\bibitem{luo} Luo M X et. al. 2015 Generations of N-atom GHZ state and 2natom W state assisted by quantum dots in optical microcavities Quantum Inf. Process. 14 116.
 \bibitem{xu} Xu, Y., Guo, Q., Si, B., Cheng, L. Y., Wang, H. F., \& Zhang, S. (2014). Generation of multi-photon Greenberger–Horne–Zeilinger states and cluster states through a quantum-dot spin in optical microcavity. Optics Communications, 313, 294-298.
\bibitem{lodahl} P. Lodahl, A. P. Mosk, and A. Lagendijk, Spatial Quantum Correlations in Multiple Scattered Light, Phys. Rev. Lett. 95 , 173901 (2005).

\bibitem{ursin} Ursin, R., Tiefenbacher, F., Schmitt-Manderbach, T., Weier, H., Scheidl, T., Lindenthal, M., ...\& \"Omer, B. (2007). Entanglement-based quantum communication over 144 km. Nature physics, 3(7), 481.
\bibitem{gvid} G. Vidal and R. F. Werner,  Computable measure of entanglement,  Phys. Rev. A65, 032314 (2002).
\bibitem{pg} P. G. Kwiat, S. Barraza-Lopez, A. Stefanov and N. Gisin,  Experimental entanglement distillation and `hidden’ non-locality,  Nature 409, 1014–1017 (2001).
\bibitem{run}  P. Rungta, V. Bu\v{z}ek, C. M. Caves, M. Hillery and G. J. Milburn,  Universal state inversion and concurrence in arbitrary dimensions,  Phys. Rev. A64, 042315 (2001).

\bibitem{aspelmeyer} M. Aspelmeyer, H. R. Bohm, T. Gyatso, T. Jennewein, R. Kaltenbaek, M. Lindenthal, G. Molina-Terriza, A. Poppe, K. Resch, M. Taraba, R. Ursin, P. Walther, and A.
Zeilinger, Long distance free-space distribution of quantum entanglement, Science 301 , 621 (2003).

\bibitem{altewischer} E. Altewischer, M. P. van Exter, and P. J. Woerdman, Plasmon-assisted transmission of entangled photons, Nature 418 , 304 (2002).
\bibitem{puentes1} G. Puentes, A. Aiello, D. Voigt, and J. P. Woerdman, Entangled mixed-state generation by twin-photon scattering Phys. Rev. A  75, 032319 (2007). 
\bibitem{dennis} Dennis Ratzel, Martin Wilkens, and Ralf Menzel Effect of polarization entanglement in photon-photon scattering, PHYSICAL REVIEW A 95, 012101 (2017).
\bibitem{aiello} A. Aiello and J. P. Woerdman, Intrinsic entanglement degradation by multimode detection, Phys. Rev. A 70 , 023808 (2004).
\bibitem{van1} J. L. van Velsen and C. W. J. Beenakker, Transition from pure to mixed-state entanglement by random scattering, Phys. Rev. A 70 , 032325 (2004).
\bibitem{jens}Eltschka, Christopher, and Jens Siewert. ``Entanglement of three-qubit Greenberger-Horne-Zeilingerâ symmetric states." Physical review letters 108.2 (2012): 020502.
\bibitem{puentes}G.Puentes, D.voigt, A. Aiello and J.P. Woerdman, Opt. Lett. 31,2057 (2006).
\bibitem{puentes2}G.Puentes, D.voigt, A. Aiello and J.P. Woerdman, Phys. Rev. A 75,032319 (2007).
\bibitem{bennett}C. H. Bennett, et. al., Phys. Rev. A 54,3824(1996).
\bibitem{vedral}V. Vedral et. al.,Prog. Quant. Electron. 22, 1 (1998).
\bibitem{bose}S. Bose et. al., Phys. Rev. A 61, 040101(R) (2000).
\bibitem{coffman}V. Coffman et. al., Phys. Rev. A 61, 052306 (2000); quant-ph/9907047.
\bibitem{valerie}Coffman, Valerie, Joydip Kundu, and William K. Wootters. ``Distributed entanglement." Physical Review A 61.5 (2000): 052306.
\bibitem{simon}Simon, R., N. Mukunda, and Biswadeb Dutta. ``Quantum-noise matrix for multimode systems: U (n) invariance, squeezing, and normal forms." Physical Review A 49.3 (1994): 1567.

\end{thebibliography}
\end{document}